# Viscoelastic Behavior of Rubbery Materials

C.M. Roland

*Polymer Physics Section*
*Naval Research Laboratory*
*Washington, DC*

# Contents

# CHAPTER 5

# Constitutive modeling, nonlinear behavior, and the stress-optic law

_______________________________________________________________________

The stresses accompanying material flow can be calculated using constitutive equations, which relate the displacements and mechanical loads in terms that are independent of sample dimensions and geometry; that is, are material properties. Constitutive equations are essential for quantitative descriptions of polymer rheology and processing. Chapters 1 and 3 discussed the Rouse model for unentangled polymers and the tube model for entangled chains. There is an enormous literature devoted to constitutive models for the linear viscoelastic response of polymer solutions and entangled melts [1-4]; however, nonlinear effects present additional challenges. Flow at high rates can reduce the degree of chain entanglement (Chapter 3) A broad consideration of the nonlinear flow behavior of polymers is eschewed herein, with consideration limited primarily to aspects unique to elastomers and filler-reinforced rubber.

## 5.1 Linearity and the superposition principle

The problem of analyzing viscoelastic mechanical behavior is greatly facilitated when the material exhibits linearity, which implies that no strain interferes with the response to any prior strain. The strict definition is: (i) stress and strain remain directly proportional to one another and (ii) time invariance of the mechanical response is observed. The stress resulting from an applied strain can then be described by the Boltzmann equation

$$\sigma(t) = \int_0^t E(t-u)\frac{d\varepsilon}{du}du \qquad (5.1)$$

where $E(t)$ is the stress relaxation function (for tension as written) and $\varepsilon$ is the uniaxial strain. Writing this equation in the form

$$\sigma(t) = \int_0^t \frac{dE(t-u)}{du}\big(\varepsilon(t)-\varepsilon(u)\big)du + E(t)\varepsilon(t) \qquad (5.2)$$

makes clear the idea that the stresses are additive with $\dfrac{dE(t-u)}{du}du$ representing the survival probability at the present time $t$ of a stress created at time $u$. Analogous expressions could be written for other modes of deformation such as shear.

For crosslinked rubber the strain can be defined in terms of the strain function suggested by the statistical theories of rubber elasticity (already defined in Chapter 4)

$$\varepsilon \equiv f(\lambda) = \lambda - \lambda^{-2} \qquad (5.3)$$

All polymers obey eqn (5.1) at sufficiently small strains; that is, within the bounds of linear behavior. For elastomers the limits of linearity extend over a wider range than for most materials [5]; however, adherence to eqn (5.1) is usually not evident from a single experiment [6]. Beyond the maximum strain for which strict linear viscoelasticity obtains, an integral constitutive equation, known as the K-BKZ equation[*] [7,8] can be employed

$$\sigma(t) = \int_0^t [E(t-u, \varepsilon(t-u))] \frac{d\varepsilon}{du} du \qquad (5.4)$$

The application of this integral is limited by the strain dependence in the kernel. If the relaxation behavior *per se* is independent of strain, the effects of time and of strain are uncoupled. Such "time invariance" permits the simplification

$$\sigma(t) = \int_0^t [E(t-u) g(\varepsilon)] \frac{d\varepsilon}{du} du \qquad (5.5)$$

In eqn (5.5) *g(ε)*, known as a damping function, accounts for the fact that the stress on the material does not remain directly proportional to strain during the course of the deformation; thus, the product *g(ε)ε* quantifies the nonlinear elasticity and *g(ε)* is unity for linear behavior. **Figure 33** and **Figure 34** respectively show relaxation data illustrating this separability of the strain and time dependences for a synthetic NR at different shear strains and for NR at three different modes of deformation [9].

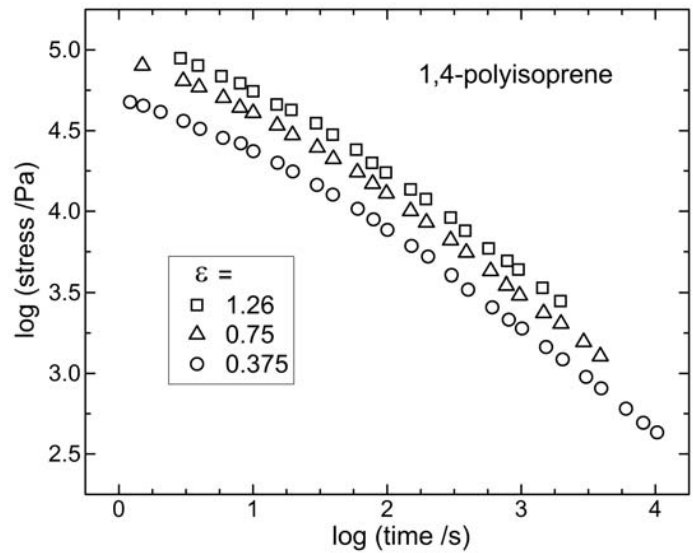

**Figure 1.** Stress relaxation of PI at the indicated shear strains [9]. Separability of time and strain is reflected in the parallel nature of the curves.

For a simple step strain the K-BKZ formulation gives

$$\sigma(\varepsilon, t) = E(t) g(\varepsilon) \varepsilon \qquad (5.6)$$

For continuous straining the integral in eqn (5.5) is approximated as a series of steps, each described by eqn (5.6). Within the limits for which this separability is valid, the integral constitutive equation, eqn (5.5), enables prediction of the stresses for arbitrary strain histories. The calculation requires knowledge of the stress relaxation function and the strain-dependence of the modulus, both determined by separate experiments. This phenomenology has often been applied to the rheology of polymer melts, with generally good results for strain-increasing deformations [10]. It can also be extended to the normal stresses, which are usually measured in shear

$$N_1 = G(t)g(\gamma)\gamma^2 \qquad (5.7)$$

were $N_1$ is the primary normal stress, acting perpendicular to the strain direction. Combining eqn (5.6) for shear and eqn (5.7) gives the Lodge-Meissner relation [11][*]

$$N_1 / \sigma = \gamma \qquad (5.8)$$

for stress relaxation after a step strain; that is, $\psi_1(t) = G(t)$.

---

[*] Equation (5.5) is known as the K-BKZ equation after the coauthors of ref. [7] and A. Kaye, who in 1962 published equivalent ideas in an obscure note to the College of Aeronautics in Cranfield, UK.

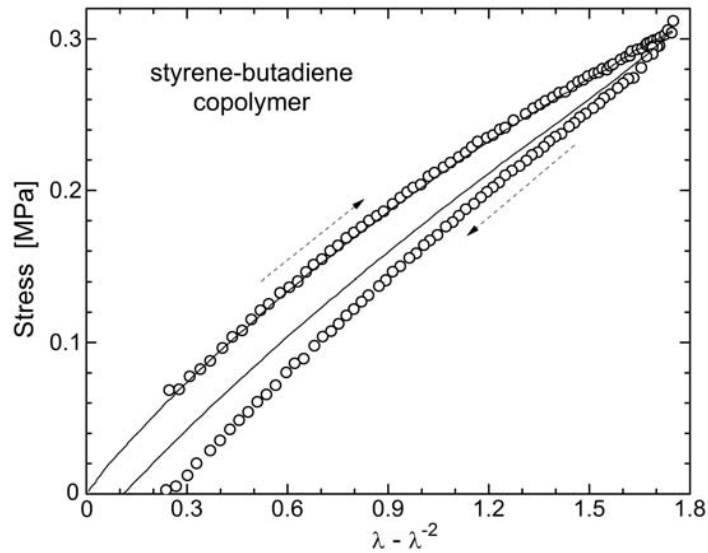

**Figure 2.** Ratio of engineering stress to strain for NR [9]. To a good approximation the stress relaxation behavior is independent of the mode of deformation.

The accuracy of K-BKZ predictions can be improved with the addition of correction terms to account for other mechanisms [13,14,15]. With the "independent alignment approximation"[†] the Doi-Edwards model of polymer dynamics yields the K-BKZ form [16]. The advantage of molecular theories is the possibility of providing a connection of the

---

[*] Much earlier Rivlin derived an equivalent relation applicable quite generally to highly elastic materials [12].

[†] The *independent alignment approximation* of the Doi-Edwards tube model simplifies the problem of calculating the coupled motion of different parts of a polymer chain by ignoring the initial stretch-retraction of the chain following a deformation. By assuming the tube does not retract, the location of a segment within the tube does not change, so that a single segment description can be used for the chain orientation.

structure to the rheological properties. The Doi-Edwards damping function for shear can be approximated as

$$g(\gamma) = \left(1 + \frac{\xi}{3}\gamma^2\right)^{-1} \qquad (5.9)$$

with $\xi$ taken as empirical parameter having a value between 0 and 0.6 [17,18]. $\xi = 0.43$ gives results approximated by an alternative expression derived from the tube model with constraint release [19].

Absent mechanisms such as strain crystallization or the finite extension of the chains, rubbery materials usually strain-soften; thus, generally the damping function decreases with strain. For elastomers in uniaxial extension the form of $g(\varepsilon)$ should be consistent with the equilibrium elastic behavior (Chapter 4). An empirical equation used to describe uniaxial extension measurements is the Mooney-Rivlin relation [20,21]

$$\sigma = f(\lambda)[c_1 + c_2 / \lambda] \qquad (5.10)$$

in which $c_1$ and $c_2$ are material constants and $f(\lambda)$ is given by eqn (5.3). The damping function can then be expressed as [22]

$$g(\varepsilon) = \left(1 + \frac{c_2}{c_1}\lambda^{-1}\right) \qquad (5.11)$$

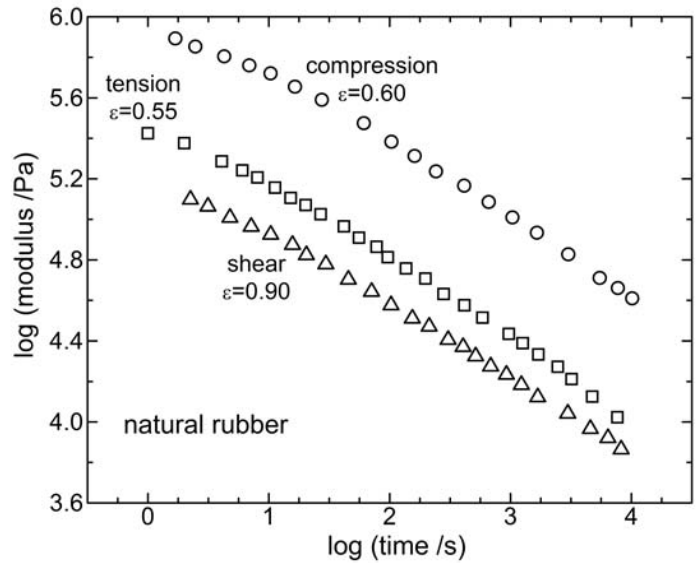

**Figure 3.** Reversing stress-strain curve for an SBR network (circles), along with the curve calculated using eqn (5.5) with the assumption of a reversible damping function. Experiments done at room temperature with a strain rate = 0.08 $s^{-1}$. Data from ref. [24].

The Mooney-Rivlin equation was derived for elastic equilibrium but can be applied to unrelaxed mechanical stresses, with a larger value of $c_2$ obtained [23].

**Figure 35** illustrates the use of eqn (5.4) to predict the stresses for a peroxide-crosslinked SBR stretched continuously at a constant velocity [24]. The modulus and stress relaxation function are both determined by separate experiments, then using these measured strain- and rate-dependences, the stress for the imposed strain history is calculated with the assumption that $g(\varepsilon)$ depends only on the instantaneous strain, a condition known as reversible damping.[*] As seen in the figure, the measured and calculated tensile stresses are equivalent.

## 5.2 Internal stress and optical birefringence

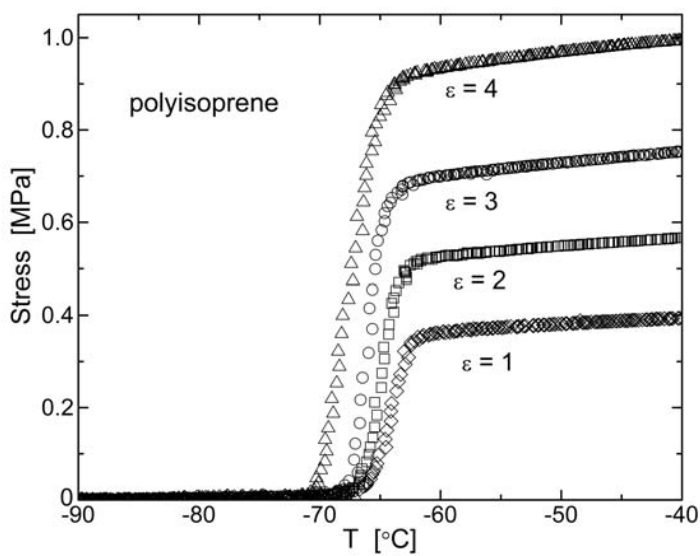

**Figure 4.** Stress measured in a crosslinked 1,4-polyisoprene during heating at 1 K/min. The sample was initially stretched at RT to the indicated tensile strain and then quenched below $T_g$. Data from ref. [25].

Constitutive models describe stresses that result from the imposition of external constraints and can be measured. However, there can be "internal stresses" within solids and very viscous liquids, which are independent of any exterior forces. These are also known as residual stresses, since a common origin is "frozen" orientation that arose during processing of a material subsequently cooled to a non-equilibrium glassy or crystalline state. **Figure 36** illustrates internal stresses in a PI network [25], which formed when the elastomer was stretched and then quenched below its glass transition temperature (= -65°C). In the glassy state no stress is measured; however, upon warming through $T_g$, the material exhibits a "memory effect", reflecting the presence of the internal stresses. Another example stresses is found in double network elastomers, which have orientation stabilized by crosslinking in the strained state (Chapter 4). Since internal stresses exert no pressure on a transducer, they cannot be measured by mechanical instruments, nor can they be described using conventional constitutive equations. Optical probes sensitive to stress are used for their characterization, with the most important being birefringence of visible light.

---

[*] Reversible damping attempts to account for any strain-induced changes in the material that do not recover on the time scale of the experiment.

### 5.2.1 Stress optic law and chain orientation.

The effect of light on matter is to polarize the material, with the induced polarization expressible as a power series in the applied electric field

$$P = \chi_1 E + \chi_2 E^2 + \chi_3 E^3 + ... \tag{5.12}$$

where the electric susceptibility $\chi$ may be a tensor. Except for intense laser sources, the linear term is sufficient to describe the response[*]. The refractive index is then independent of the light intensity with the Lorentz-Lorenz equation giving

$$\frac{n^2 - 1}{n^2 + 2} = \frac{4\pi}{3} N \chi_1 \tag{5.13}$$

in which $N$ is the number density of molecules (chain segments) having polarizability $\chi_1$. Classically, the material is considered a continuous dielectric with certain symmetry properties. The molecular origin of the polarizability is the dispersibility of the electron cloud, and since the outer electrons are more easily displaced along their bond axis than transversely, this polarizability is inherently anisotropic; however, the average over many randomly arranged molecules yields a symmetric refractive index. When deformation induces orientation, the principal components of the refractive index, $n_x$, $n_y$ and $n_z$, become non-equal, causing birefringence, $\Delta n$, of light propagating through the material. Birefringence ("double refraction") refers to a difference between the refractive indices within the material for the two transverse polarizations of propagating light. This birefringence can be used to characterize the deformation according to the stress optic law[†], which states that the anisotropy of the refractive index is proportional to the difference in the principal stresses. For uniaxial stress in the $x$-direction, $(|\hat{\sigma}_x| > 0; \ \hat{\sigma}_y = \hat{\sigma}_z = 0$, where the caret signifies true stress[‡])

$$\Delta n \equiv n_x - n_y = n_x - n_z = C \hat{\sigma}_x \tag{5.14}$$

---

[*] The quadratic term in eqn (5.12), which vanishes for molecules possessing an inversion center, is responsible for hyper-Raman scattering and optical rectification, while $\chi_3$, describing events involving three simultaneous photons, gives rise to stimulated Raman scattering, coherent anti-Stokes Raman scattering, and the optical Kerr effect.

[†] The stress optic law is sometimes referred to as Brewster's Law, after the Scottish physicist who observed compression-induced birefringence of glass in the early 19[th] century. The subsequent discovery of the proportionality between stress and birefringence was by Maxwell. Brewster's Law also refers to an equation for the angle of incidence at which a reflected light beam will be polarized. "Brewster" is a unit of birefringence equal to 1 TPa[-1].

[‡] The true or Cauchy stress is the force relative to the instantaneous area. For uniaxial strain, $\hat{\sigma} = [1 + \varepsilon] \sigma$; for shear $\hat{\sigma} = \sigma$.

$C$ is the stress-optical coefficient, a material constant determined by the chemical structure of the repeat unit. Equation (5.14) has long been exploited to quantify stresses and their spatial distribution in glasses, plastics, and other transparent materials, wherein the relation is known as the photoelastic effect. The magnitude of $\Delta n$ also reflects internal stresses, since it arises from the anisotropy of the polarizability and thus only requires molecular ordering, not external forces. An example is double network elastomers (Chapter 4), which as mentioned above, are internally stressed. As shown in **Figure 37** (see also Figure 18 in Chapter 4), this leads to birefringence notwithstanding the absence of external stress or strain.

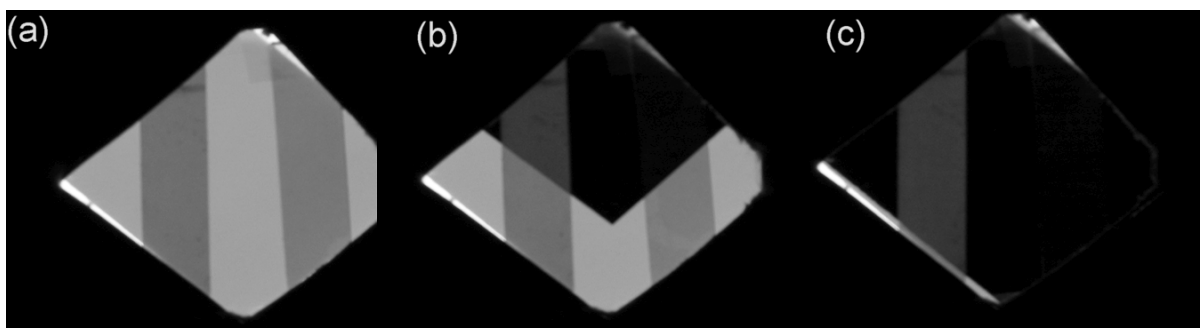

**Figure 5.** (a) Natural rubber double network (on left) and conventional NR elastomer (on right), both in a state of mechanical equilibrium ($\sigma = 0$); nevertheless, (b and c), through crossed polarizers the double network transmits light, due to its inherent orientation. Since the external stress is zero, this is a violation of the stress optic law, eqn (5.14).

The stress optic law has been widely applied to deformed rubber. The repeat units of chain molecules are optically anisotropic, so macroscopic orientation leads to birefringence conforming to eqn (5.14), as first reported by Kuhn and Grün [26] proportionality of $\Delta n$ and the stress of a deformed elastomer. Assuming phantom network behavior (Chapter 4), their result is

$$\Delta n = \left[ \frac{2\pi \Delta \alpha \left( n^2 + 2 \right)^2}{45 kTn} \right] \hat{\sigma} \tag{5.15}$$

where $\Delta \alpha$ is the optical anisotropy of the chain segment, $n$ is the refractive index in the absence of orientation, and the bracketed factor defines the stress optical coefficient. $C$ is independent of crosslink density or molecular weight and can be used to determine the difference in parallel and perpendicular polarizabilities of the segment. **Figure 38** shows the birefringence measured for two crosslinked rubbers subjected to both tension and compression [27]. A linear fit through the origin ($\varepsilon = 0$) represents the data well, in accord with eqn (5.15) for phantom networks; however, more realistic elasticity models predict a small difference in $C$ for different modes of deformation, so that the slope in Figure 38 would change on passing through the origin.

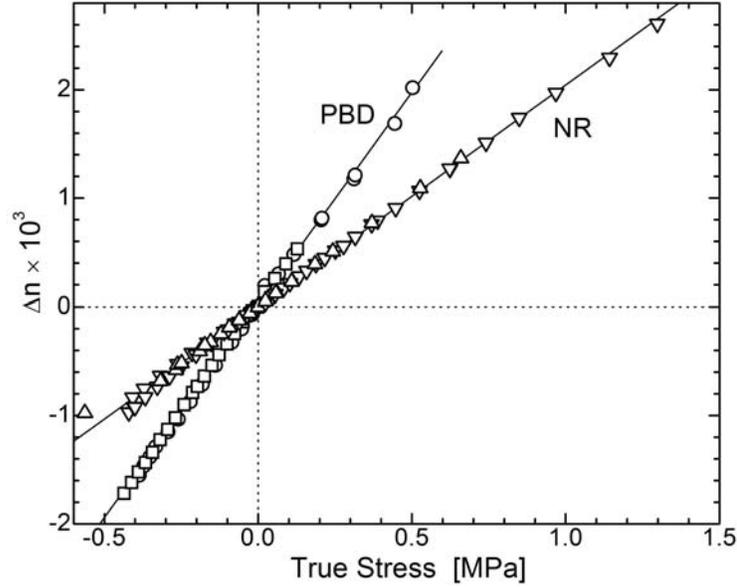

**Figure 6.** Birefringence as a function of true stress for the cross-linked 1,4-polybutadiene and natural rubber subjected to tension and compression; the slopes yield 3.6 and 2.0 GPa$^{-1}$ for the respective stress optical coefficients. Different symbols represent different test specimens. Data from ref. [27].

The measured birefringence represents the average value of all segments, each having a probability of alignment at a given angle to the reference direction described. This can be described by an orientational distribution function, which for uniaxial orientation of linear (axially symmetric) chains can be expressed as a series

$$f(\theta) = \sum_{j=0}^{\infty} a_j P_j(\cos\theta) \qquad (5.16)$$

where $P_j(\cos\theta)$ is the $j^{th}$ Legendre polynomial* and $\theta$ is the angle the long axis of the segment makes with the reference axis. It is a property of Legendre polynomials that the weighting factor in eqn (5.16) is just the mean value times a constant factor [28]

$$f(\theta) = \sum_{j=0}^{\infty} \frac{2j+1}{2} \left\langle P_j(\cos\theta) \right\rangle P_j(\cos\theta) \qquad (5.17)$$

The relevant Legendre function will depend on the experimental variable (the order of its tensor), although X-ray scattering can in principle yield the complete orientation function, enabling determination of all orders. Birefringence is related to the second Legendre polynomial

$$f_2(\theta) = \frac{\Delta n}{\Delta n_0} = \frac{3\left\langle \cos^2\theta \right\rangle - 1}{2} \qquad (5.18)$$

---

* Legendre polynomials are given by the formula $f_n = \dfrac{1}{2^n n!} \dfrac{d^n}{dx^n}\left[ \left(x^2 - 1\right)^n \right]$ , $n = 0, 1, 2, \ldots$

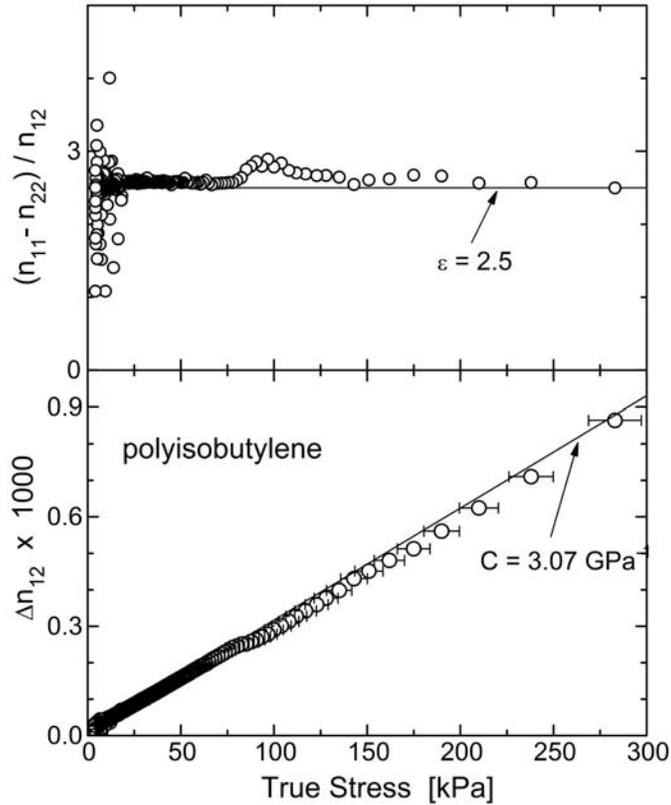

**Figure 7.** (bottom) Birefringence *versus* shear stress for uncrosslinked PIB following a step shear strain = 2.5. Solid curve is the linear fitting yielding the indicated stress optical coefficient. (top) Ratio of refractive index components (1 is the flow direction and 2 the direction of the gradient). is constant and equal to the shear strain, in accord with eqn (5.8). Data from ref. [30].

Equation (5.18) is known as the Hermans orientation function.* The quantity $\Delta n_0$, proportional to $\Delta\alpha$, is the magnitude of the birefringence for complete alignment, whereby $f_2(\theta) = 1$ for perfect orientation. For random orientation the average is over all angles

$$\left\langle \cos^2\theta \right\rangle = \int_0^\pi \cos^2\theta \left[\frac{1}{2}\sin\theta\right] d\theta = \frac{1}{3} \tag{5.19}$$

where the bracketed quantity is the weighting function for $\theta$; thus $f_2(\theta) = 0$.

The Hermans orientation function can also be determined from the dichroic ratio, defined as the ratio of the infrared absorbance for radiation polarized parallel and perpendicular to the stretching direction [29]

$$f_2 = \frac{(R-1)(R_0+2)}{(R+2)(R_0-1)} \tag{5.20}$$

---

* Named for the Dutch scientist P.H. Hermans, eqn (5.18) is known in the liquid crystal literature as the order parameter.

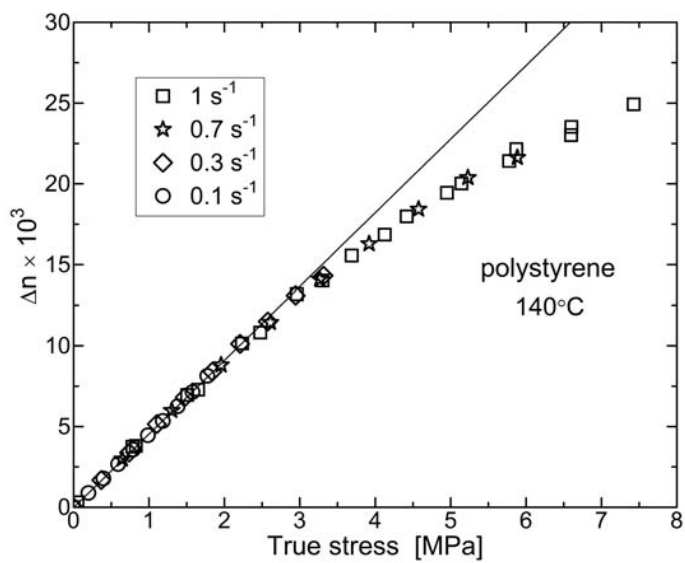

**Figure 8.** Birefringence *versus* true stress for a PS melt (Mw = 206 kg/mol) during uniaxial extension at the indicated strain rates. The data superpose with deviation from the stress optic law observed for $\hat{\sigma} > 2.2$ MPa. Data from ref. [33].

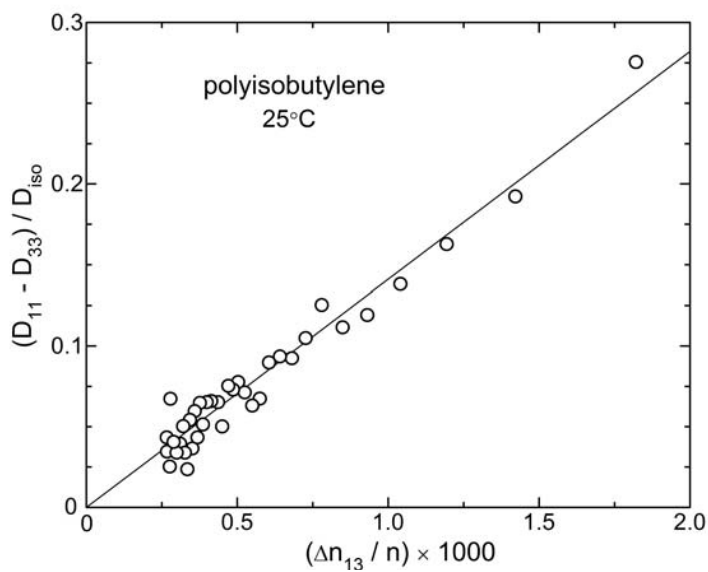

**Figure 9.** Normalized difference in thermal diffusivity *versus* normalized birefringence for polyisobutylene ($M_w$ = 85 kg/mol) after a step shear strain equal to eight. The solid line through the data suggests a proportionality between these quantities. Data from ref. [36].

$R_0$ is the dichroic ratio for perfect orientation, $R_0 = 2\cot^2\varphi$, with $\varphi$ the angle between the chain axis and the transition moment vector of the vibration. The Hermans orientation function can also be obtained from fluorescence polarization and from analysis of the amorphous halo in X-ray scattering.

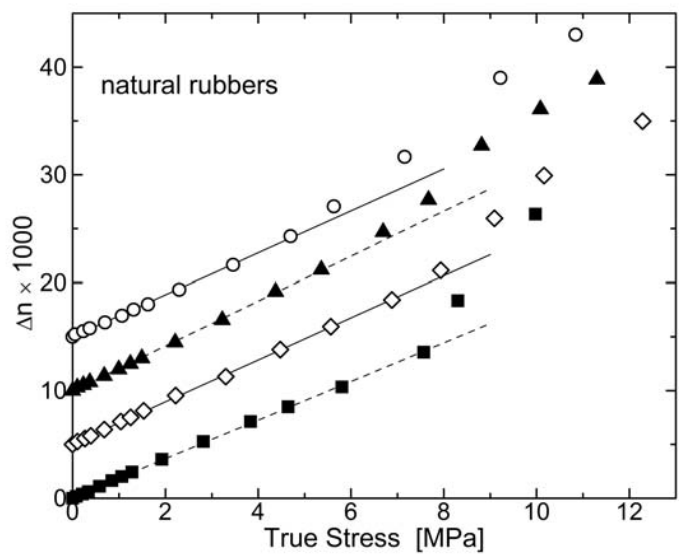

**Figure 10.** Birefringence measured at room temperature for natural rubber elastomers, in order of increasing purity of the material: guayule NR (circles); "ribbed smoked sheet" grade of NR (triangles); "SML" grade NR (diamonds); deproteinized NR (squares). The solid lines are fits to the data at lower strains, yielding a mean $C = 1.9 \pm 0.1$ GPa$^{-1}$. An excess $\Delta n$ appears at higher strains due to strain-induced crystallization. The data, from ref. [38], are shifted vertically for clarity.

Equations (5.15) and (5.18) can be used to determine the anisotropy of, not only the network chains in an elastomer, but also deformed or flowing polymer melts and solutions. **Figure 39** [30] shows that conformance to the stress optic law is maintained during stress relaxation of an uncrosslinked PIB for both the shear and normal components. The upper half of this figure also confirms the Lodge-Meissner relation, eqn (5.8). The stress optical behavior is independent of strain rate [31,32,33], although as seen in data for stretched PS melts (**Figure 40** [33]), the stress optical coefficient decreases in the nonlinear regime. The latter effect increases in prominence for more polydisperse materials, probably related to the varying fraction of chains oriented by the applied field [34]. Since orientation of rubber also affects heat conduction [35], there is a corresponding proportionality, known as the "stress-thermal rule", between stress and the anisotropy of the thermal conductivity [36]. This in turn causes the latter to be proportional to the birefringence (**Figure 41**) [30,36].

The Hermans orientation function quantifies the mean orientation of the segments, and it is this microscopic or molecular strain that $f_2(\theta)$ connects to $\Delta n$. Thus, any changes in the molecular polarizability *per se* can cause birefringence unrelated to molecular orientation, and likewise, orientation due to internal stress will contribute to $\Delta n$. Both these situations give rise to deviations from the stress optic law. An example is hydrodynamic drag forces [37], constituting the fluid resistance to motion. These can orient the polymer chains but do not affect the measured stress; thus, their contribution to $\Delta n$ is not accounted for by the stress optic law. Another example is when the material itself changes, such as by strain-induced crystallization. The intrinsic birefringence of crystals can be very high, which in combination with a contribution from "form birefringence" (refraction at the interface between crystalline and amorphous domains) makes measurements $\Delta n$ very sensitive to the presence of crystallinity. This is illustrated in **Figure 42** showing $\Delta n$ as a function of stretch for various natural rubber networks [38]. Departures from the stress optic law are observed at strains for which other techniques give no indication of crystallization.

### 5.2.2. Birefringence in the glass transition zone

More interesting, but of less practical value, are departures from proportionality between $\sigma$ and $\Delta n$ that arise when there is a disconnect between stress and the chain orientation. One example is during deformation at frequencies high relative to the rate at which the chains alter their configuration through backbone bond rotation. In this circumstance the material responds by stretching and bending of these bonds, which generate stresses not described by eqn (5.15). Nevertheless, birefringence measured well within the glassy zone of the viscoelastic spectrum is still proportional to the stress, but with a different value of the stress optical coefficient.

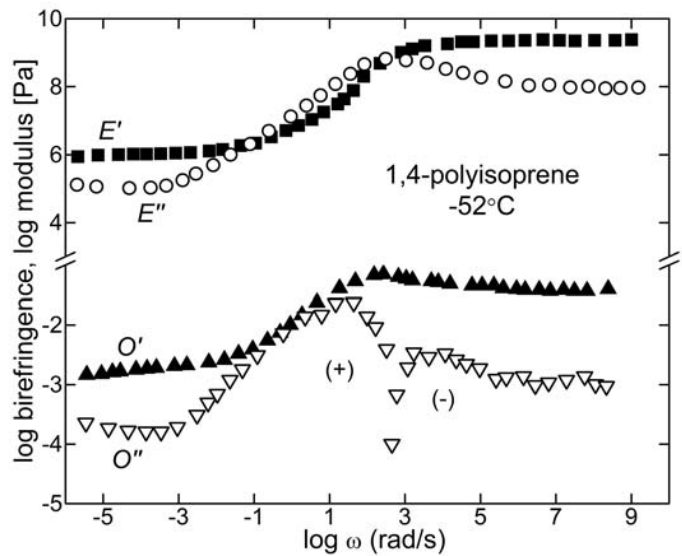

**Figure 11.** Time-temperature master curves of the dynamic moduli (upper curves) and strain optical coefficients (lower curves) for PI. There is a change in sign of $O''(\omega)$ at log $\omega = 2.5$. Data from refs. [39].

This difference in the stress optical coefficients causes breakdown of the stress optic law in the softening regime of the viscoelastic spectrum, where both chain orientation and bond distortion mechanisms are operative. Experimental measurements show that $\Delta n^*$ and $\sigma^*$ have very different frequency dependences, as observed in the glass transition zone of master

curves[*] for PI in **Figure 43** [39]. To analyze this region eqn (5.14) is rewritten in complex form

$$\Delta n^* = C(\omega)\ \sigma^*(\omega) \tag{5.21}$$

Dividing by the strain gives the strain-optical coefficient

$$O^*(\omega) = C(\omega)\ E^*(\omega) \tag{5.22}$$

The birefringence measured in the softening zone can be assumed to have two independent components [40,41,42]

$$\Delta n^*(\omega) = C_G\hat{\sigma}_G^*(\omega) + C_R\hat{\sigma}_R^*(\omega) \tag{5.23}$$

In these equations $\hat{\sigma}_G^*$, $\hat{\sigma}_R^*$ and $C_G$, $C_R$ are the respective complex stresses and stress optical coefficients for the "glassy" (chain distortion) and "rubbery" (chain orientation) components, the latter underlying rubber elasticity and the Rouse dynamics. Equation (5.23) follows from the assumption that the corresponding stresses are additive

$$\hat{\sigma}^*(\omega) = \hat{\sigma}_G^*(\omega) + \hat{\sigma}_R^*(\omega) \tag{5.24}$$

This means that stresses within the same chain segment arise independently and are completely decoupled. The expression for the complex strain optical coefficient is then

$$O^*(\omega) = C_G E_G^*(\omega) + C_R E_R^*(\omega) \tag{5.25}$$

with $E_G^*$ and $E_R^*$ the respective components of the dynamic modulus

$$E^*(\omega) = E_G^*(\omega) + E_R^*(\omega) \tag{5.26}$$

To implement this approach beyond a trivial, multi-parameter curve-fitting exercise, some definition must be adopted for the stress optical coefficients. Read assumed [40]

$$C_R = \lim_{\omega \to 0}\frac{O'(\omega)}{E'(\omega)} = \lim_{\omega \to 0}\frac{O''(\omega)}{E''(\omega)} \tag{5.27}$$

and

$$C_G = \lim_{\omega \to \infty}\frac{O'(\omega)}{E'(\omega)} \tag{5.28}$$

Inoue and coworkers [41,42] employed a different definition for the glassy coefficient

$$C_G = \lim_{\omega \to \infty}\frac{O''(\omega)}{E''(\omega)} \tag{5.29}$$

These equations have been applied to dynamic birefringence data on various polymers [40,41,42,43]. The analyses yield predictions for the frequency-dependent contributions of the high and low frequency components. Two questions are apparent:

---

[*] As discussed in Chapter 6 time-temperature superpositioning is invalid in the glass transition zone because of the different temperature-dependences of the orientation and distortion mechanisms; therefore, the shape of the master curve in this region is not quantitative.

### 5.2.2.1. Are the assumptions of this approach (eqns (5.23) through (5.29)) correct and/or verified by their application?

The only quantities not experimentally determined are the (hypothetical) components of the modulus, which from eqns (5.23) and (5.26) are given by

$$E_G^*(\omega) = \frac{O^*(\omega) - C_R E^*(\omega)}{C_G - C_R} \tag{5.30}$$

and

$$E_R^*(\omega) = \frac{O^*(\omega) - C_G E^*(\omega)}{C_R - C_G} \tag{5.31}$$

Dynamic birefringence data can be described equally well using either eqns (5.28) or (5.29), but with different values obtained for the high and low frequency components. This means that the approach of Read [40] and Inoue et al. [41,42,43] is ambiguous and thus not corroborated by successful fitting of experimental data.

The premise of the method is that the stresses are additive. This is consistent with the virial stress formulation, used in molecular dynamics simulations, in which the atomic virial stresses[*], arising from the kinetic and potential energies, are summed to obtain the (continuum) Cauchy stress [39,44]. However, simulations can also be analyzed using alternative conservation laws, for example based on summation of the local strains [45]. Experimentally it is known that the total strain during creep flow of linear viscoelastic materials is equal to the sum of the recoverable strain and the viscous flow; in terms of the shear compliance, this strain additivity is expressed as

$$J(t) = J_r(t) + t / \eta_0 \tag{5.32}$$

where $J_r$ is the recoverable creep compliance. The validity of eqn (5.32) is confirmed by creep-recovery experiments on polymer melts, in which the permanent deformation measured after steady state flow is equal to $t/\eta_0$ times the creep stress [46]. Equation (5.32) indicates that at least for terminal flow, stresses (elastic and viscous) are coupled rather than additive.

Analysis of birefringence in the softening zone can be carried out using the assumption that the strains are additive [47]; thus,

$$\varepsilon = \varepsilon_G + \varepsilon_R \tag{5.33}$$

In terms of the compliances, $D^*(\omega) = 1/E^*(\omega)$,

$$D^*(\omega) = D_G^*(\omega) + D_R^*(\omega) \tag{5.34}$$

---

[*] The term "virial" refers to the configurational part of the pressure, with the virial stress given by $\sigma_{12} = V^{-1} \sum_i \left( \frac{1}{2} \sum_j \left( r_{1,i} - r_{1,j} \right) F_{2,ij} - m_i v_{1,i} v_{2,i} \right)$ where the index (1,2) denotes directions in a three-dimensional rectilinear coordinate system, and the sums are over all $i$ atoms in volume $V$, interacting through a force $F_{ij}$ with each $j^{th}$ neighbor. The last term is the thermal kinetic energy. The virial pressure for an ideal gas is zero.

with the stress optic law given by

$$O_G^*(\omega)^{-1} = \frac{D_G^*(\omega)}{C_G} + \frac{D_R^*(\omega)}{C_R} \qquad (5.35)$$

The assumption of strain additivity gives for the glassy component [48]

$$D_G^*(\omega) = \frac{C_R / O^*(\omega) - D^*(\omega)}{C_R / C_G - 1} \qquad (5.36)$$

and for the rubber component

$$D_R^*(\omega) = \frac{C_G / O^*(\omega) - D^*(\omega)}{C_G / C_R - 1} \qquad (5.37)$$

Note that $C_R$ and $C_G$ are no longer defined by eqns (5.27), (5.28) , or (5.29); without additional assumptions, they are free parameters adjusted to give agreement between the calculated and experimental birefringence data.

### 5.2.2.2. Are the results of eqns (5.23) through (5.29) reasonable?

The most important aspect of such analyses is what they reveal about the molecular mechanism underlying the viscoelastic behavior. Ostensibly the relative contribution of the chain segment reorientation and the distortional modes to the various parts of the viscoelastic spectrum can be quantified. In **Figure 44** these contributions, as reflected in the magnitudes of rubbery and glassy dynamic moduli, are plotted versus frequency. Of course the rubbery modes dominate at the lowest frequencies, while glassy component governs the behavior at the highest frequencies. More interesting is the relative contribution in the softening zone in Fig. 9 (*ca.* -1 < log $\omega$ (rad/s) < 3). The assumption of stress additivity leads to the inference that the glassy modes are the dominant relaxation mechanism throughout the transition region, including at frequencies over which the loss component of the birefringence, $O''(\omega)$, changes sign. When strain additivity is assumed, however, the relative contribution of the glassy component is significantly smaller throughout most of the softening zone. And the main contribution changes from the glassy to the rubbery component at the same frequencies for which $O''(\omega)$ changes sign. Thus, the different assumptions used in interpreting dynamic stress birefringence measurements lead to diametric conclusions.

One difficulty with results of this type is the reliance on master curves (e.g., Figure 43), obtained by time-temperature shifting. Although this is necessary in order to obtain the entire viscoelastic spectrum, superpositioning breaks down for polymers in the glass transition zone, as discussed in Chapter 6. While the interpretation of dynamic birefringence can potentially yield useful insights, the interplay of forces in a dynamically correlated system such as polymer melts may be too complicated for an analysis in terms of distinct "rubbery" and "glassy" components. It remains unclear how these putative entities interact, so that rote application of the stress optic law to behavior in the softening zone is dubious.

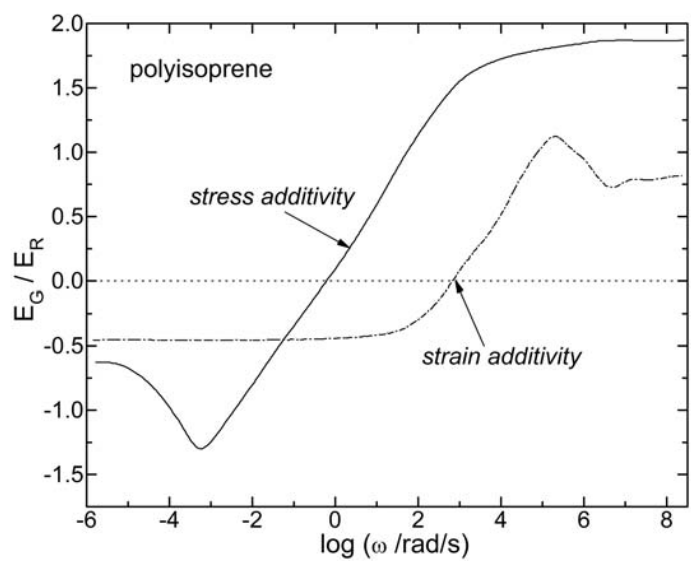

**Figure 12.** The relative contribution of the distortional and orientational components of the stress birefringence, as reflected in the magnitude of the respective glassy and rubbery moduli, determined from the data in **Figure 43** using eqn (5.25) (solid line) or eqn (5.34) (broken line). The horizontal dotted line denotes equal contribution from the two components.

### 5.2.3. Orientational coupling (nematic interaction)

Since the mechanism of birefringence arises on the level of the chemical bonds, whereas rubbery stresses involve orientation of subchains involving numerous backbone bonds, another departure from proportionality between stress and birefringence can result from local ordering that does not affect the stress. In the classical depiction of rubber elasticity, network chains orient independently in response to applied stresses; however, intermolecular interactions, mainly steric effects from aligned neighboring segments, can also contribute to the local orientation. For this reason unattached polymer chains acquire a stable orientation when present in an oriented network [49,50,51,52,53,54]. Similar effects are seen in polymers with bidisperse molecular weight distributions [55,56,57,58] and in miscible blends of chemically dissimilar polymers [59,60]. This orientational coupling (or nematic interaction) in polymer melts causes the relaxation times of shorter chains to approach the value of $\tau_\eta$ for the longer host chains.

Orientational coupling is only locally effective, so that short probe chains in a strained network retain their isotropic dimensions [61,62] and there is no effect on the stress. However, orientational coupling contributes to optical birefringence and can also be detected by infrared dichrosim, deuterium NMR by the quadrupole interaction, fluorescence polarization, and from the anisotropy of neutron scattering. Molecular dynamics simulations of chains interacting via a Lennard-Jone 6-12 potential also reveal nematic interactions [63,64]. A metric of the strength of the orientational coupling is the ratio of the strain of the probe chain to that of the matrix [65]. Values of this coupling coefficient for different probes in various rubbery hosts are listed in **Table 1.** In mixtures of different polymers, including bidisperse molecular weight distributions, the stress optic law breaks down because of the orientational coupling; however, it does not affect the proportionality between the stress and the total optical anisotropy from all chains, so that eqn (5.14) is maintained.

**Table 1.** Coupling coefficients for various probe/host mixtures.

| materials | comment | coupling coefficient | method | reference |
|---|---|---|---|---|
| PMPS | | 1 | $^2$H NMR | 61 |
| PBD | probe $M_w > 2$ kg/mol | $0.9 \pm 0.1$ | birefringence / IR dichroism | 51 |
| | probe $M_w > 2$ kg/mol | $0.4 \pm 0.05$ | | |
| poly(ethylene-*co*-propylene) | $M_w$ (kg/mol) = 53 – 370 | $0.45 \pm 0.05$ | birefringence / IR dichroism | 55 |
| polydiethylsiloxane (PDES) | $M_w$ (kg/mol) = 27 – 16310 – 71 | > 0 | SANS | 62 |
| modified PBD | 1 – 4% H-bonding monomer units | 0.6 – 1.0 | birefringence / IR dichroism | 58 |
| polyisoprene | small molecule probes | > 0 | $^2$H NMR | 49 |
| polyisoprene | small molecule probes | 0.4 – 0.65 | fluorescence polarization | 65 |
| PVE/PI blend | PVE $M_w$ (kg/mol) = 23 – 134; PI network | 0.13 – 0.20 | IR dichroism | 52 |
| | PVE $M_w$ (kg/mol) = 80 – 110; linear PI | 0.30 – 0.35 | birefringence | 60 |
| | PVE $M_w$ (kg/mol) = 204; linear PI | $0.35 \pm 0.05$ | birefringence / IR dichroism | 59 |
| PDMS/PDES blend | PDMS $M_w$ (kg/mol) = 15; PDES network | > 0 | $^2$H NMR | 53 |
| bead-spring model (L-J 6-12 potential) | probe chains: N = 5 – 25; host chains: N = 5000 | $0.28 \pm 0.01$ | simulations | 63 |
| | probe chains: N = 10 – 20; host chains: N = 20 – 120 | $0.35 \pm 0.05$ | simulations | 64 |

## 5.3 Reversing strain histories

The Boltzman superposition approach (eqn (5.1)) treats a continuous deformation such as constant strain rate as a sequence of incremental, discrete strains separated by a small time interval during which the chains relax. Strict linearity implies the material response to each step is independent of all other responses. As discussed above, when this condition is not met, nonlinearities can be incorporated into the definition of the strain, so that the stress remains proportional to a strain function which is factored separately from viscoelastic effects (eqn (5.5)). However, this method fails when the deformation changes sign (direction). Reversing strain histories in general give rise to anomalous behavior, as illustrated here with two examples, the Mullins effect and the optical birefringence during recovery from creep strain. These anamolies reveal the need for further development of molecular-based constitutive equations for both entangled melts and rubbery networks.

### 5.3.1. Mullins effect

A change in the sign of the strain during deformation of a rubber, such as retraction of stretched rubber, causes stresses that are lower than the corresponding values prior to the reversal [66]. The only exception to this behavior is low strain rates yielding an elastic response without accompanying structural or chemical changes. Mechanisms for mechanical hysteresis during reversing strains include: (i) Changes in chemical structure, for example rupture of network chains ("chemical stress relaxation") [67,68,69,70] or their detachment from filler particles [71,72,73]. These effects are irreversible or recover only very slowly, so that the reduction in stiffness persists throughout the deformation cycle. (ii) Crystallization induced by large deformation. Orientation decreases the conformational entropy and thus the entropy penalty associated with crystal formation. Strain crystallization in turn lowers the stress by reducing the number of elastically effective network strands and diminishing their microscopic strain[*] [74,75,76]. The reduced backfolding (crystal stem reentry) of crystals formed while oriented leads to more stable crystals, which may persist even after removal of the strain. (iii) The third origin of strain softening is viscoelasticity, a phenomenon common to all rubbers as described in Chapter 1. The response of the network chains is retarded due to their mutual interactions; thus, unless mechanical equilibrium is maintained through low strain rates, the recovery stresses are smaller.[†]

Although the term Mullins softening is commonly applied to the mechanical hysteresis of carbon black filled rubbers, this is an unfortunate use of the term. The effect is not specific to filler reinforcement, as first shown by Harwood and coworkers [77,78]. They found for cyclically stretched rubber that the retraction stresses were independent of carbon black content, as long as the strain for each compound was adjusted to give constant maximum stress. The maximum strain was reduced as the filler content increased, yielding a roughly constant degree of stress-softening. This comparable behavior of filled and gum rubbers is seen in plots (**Figure 45**) of the stress during cyclic straining, plotted as a function of the stress during the initial extension [77].

---

[*] At mechanical equilibrium strain-induced crystallization reduces the stress in accord with Le Chatelier's principle; however, by functioning as inextensible filler particles, the crystallites usually increase the transient stress.

[†] This type of mechanical hysteresis is analogous to the enthalpy changes accompanying strain-crystallization of rubber. The heat evolved during the initial stretching is greater than that absorbed during the subsequent recovery because the latter only reflects the latent heat of crystallization, while the former has a contribution from internal friction. The result is an increased temperature of strain-crystallizing rubber when cyclically strained.

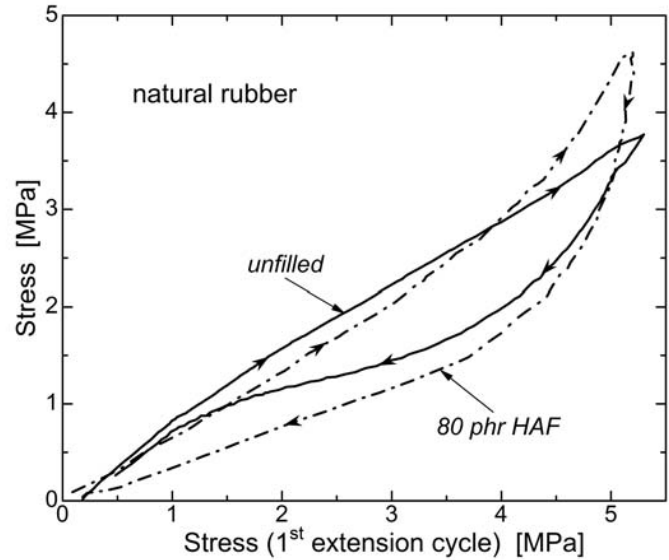

**Figure 13.** Engineering stress for networks of gum NR and NR with carbon black during the 2nd cycle of extension-retraction *versus* the stress for the initial extension. The strain rate was 0.03 s⁻¹. Data from ref. [77].

As discussed above, stresses calculated using K-BKZ superpositioning are generally in good agreement with measurements on elastomers for shear or tensile strains up to at least 100%. However, this success is limited to stress-increasing strain histories. When the sign of the strain changes, the energy dissipation is significantly underestimated. This is the Mullins effect, which refers specifically to *anomalously* small recovery stresses. The stresses during retraction are smaller than can be accounted for by the mechanisms enumerated above (bond breaking, crystallization, or viscoelastic relaxation) [77,78,79]. Thus, assessment of Mullins softening requires quantitative evaluation of the viscoelastic contribution to the hysteresis. This can be done with experiments on amorphous gum rubber, lacking hysteresis due to filler or crystallization. If the crosslink junctions are covalently bonded carbon atoms, rather than polysulfidic linkages, and strains are limited to less than *ca.* 100%, the possibility of bond rupture can be negated. Mulling softening is illustrated in **Figure 35** showing data for an SBR network; the stresses during retraction are significantly lower than the calculated values.

A simpler test employs a two-step deformation, with the stress from eqn (5.5) expressed as the sum of the stresses measured in single-step experiments

$$\sigma(t,t_1,\gamma_1,\gamma_2) = \sigma(t,\gamma_1+\gamma_2) - \sigma(t,\gamma_2) + \sigma(t-t_1,\gamma_2) \qquad (5.38)$$

where $\gamma_1$ is the shear strain applied at $t = 0$ and the total strain becomes $\gamma_1 + \gamma_2$ at $t = t_1$. As seen in data for peroxide-crosslinked NR (**Figure 46**) [80], the K-BKZ expression over-predicts the stress for $\gamma_2 < \gamma_1$ (strain reversal). This failure for reversing deformations of the (otherwise successful) Boltzmann superposition integral with time invariance implies that the calculation of the retraction stresses is in error; however, why this error becomes apparent only for reversing strains is unclear. Evidently there is an unaccounted for contribution to the viscoelastic response manifested only during recovery.

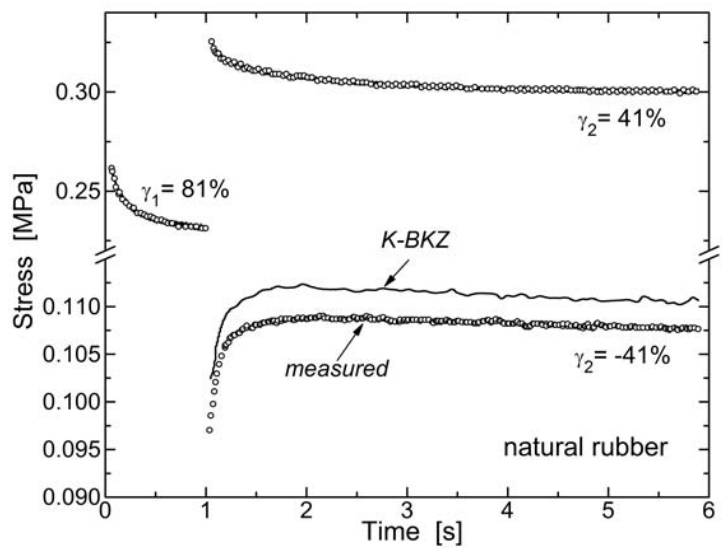

Figure 14. Stress measured for cross-linked NR subjected to double-step shear strains (circles) of the indicated magnitudes and sign. The second step was applied at $t_1 = 1$ s after the initial strain. The stresses calculated from eqn (5.38) for $t > t_1$ (solid line) are only evident for the reversing strain history ($\gamma_2 + \gamma_1 < \gamma_1$; lower curves), being masked by the experimental points for the stress-increasing deformation. Data from ref. [80]

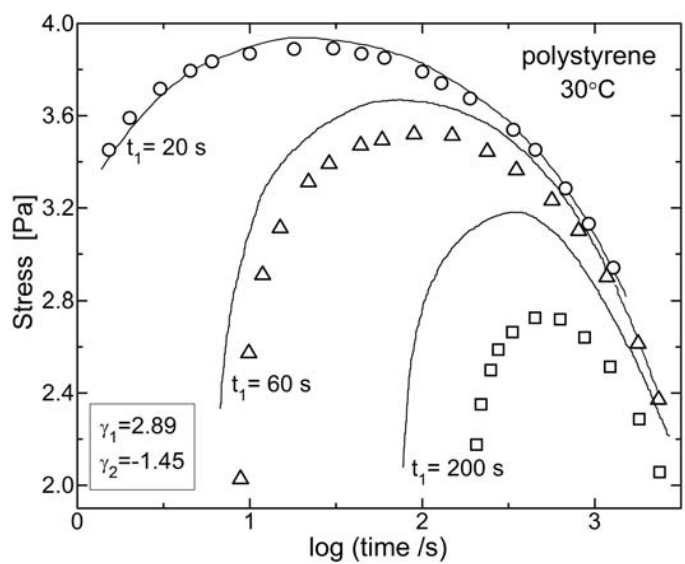

**Figure 15.** Shear stress for polystyrene ($M_w = 670$ kg/mol) solution for shear strain = 2.89 imposed at $t = 0$ and reduced to 1.44 after the indicated time intervals. Equation (5.5) (solid curves) increasingly over-estimates the stresses with increasing time between successive steps. Data from ref. [84].

This same anomalous hysteresis is also observed for polymer melts and is relevant to processing operations that involve strain reversals, such as calendering, extrusion, blow molding, etc. Constitutive models having the form of eqns (5.5) or (5.38) cannot quantitatively describe reversing deformations of uncrosslinked polymers [18,81,82,83]. This is illustrated in **Figure 47** showing data for a concentrated solution of entangled polystyrene subjected to a double-step strain [84]. Similar to the results in Figure 35 and Figure 46 for rubbery networks, the calculated stresses are higher than the measures values, with the deviation growing with increase in the time interval between application of the second strain. Of course, for sufficiently large $t_1$ the steps become independent since the initial strain fully relaxes and, in the limit $t_1 = 0$, the reversing strain experiment reduces to simple stress relaxation at a strain equal to the sum of the two steps. The discrepancy in calculations involving a change in sign of the strain for entangled polymer melts is ascribed to chain end retraction [2,9]. A chain is assumed to respond affinely to deformation, but since entanglements do not constrain motion along the chain contour, the equilibrium contour length is recovered when the chain retracts (according to the tube model strain orients the chain segments without stretching them). This nonaffine motion is very rapid, involving conformational transitions of independent chain segments. (It serves as the mechanism responsible for shear thinning behavior of polymer melts [2].) If too fast to be included in the measurement ("visco-anelasticity" [18]), the reduction in stress is omitted from the memory kernel of the constitutive equation, with the error in the calculation only becoming evident for strain reversals.

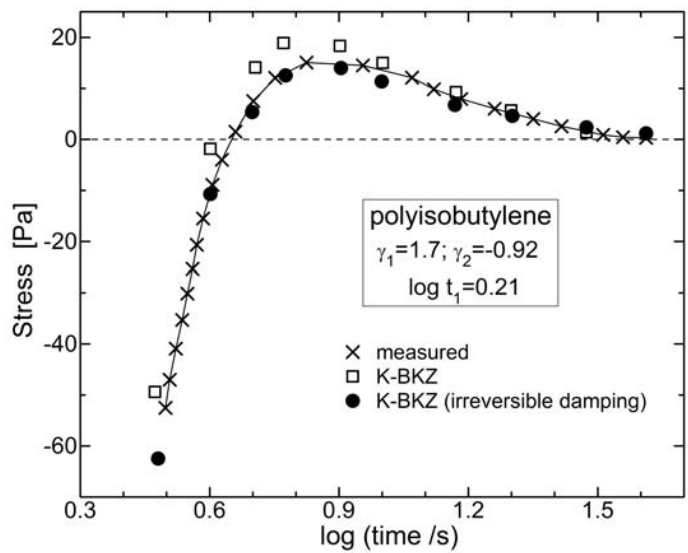

Figure 16. Stress measured for a PIB melt subjected to a shear strain of 1.77 at t = 0 that was reduced to 0.85 at t = 1.62 s (crosses), along with the stresses calculated using eqn (5.5) assuming reversible (squares) and irreversible (circles,) damping. Data from ref. [10].

There have been various attempts to improve on the K-BKZ description of reversing strains [85,86,87,88]. In the Doi-Edwards model this failing has been addressed by removing the independent alignment assumption [16,18,84]. Wagner [81,89] introduced irreversibility into the calculation by requiring that the damping function retain its minimum value, attained at the highest strain. The underlying idea is that chain end retraction causes a near instantaneous loss of stress that is only sensed during a strain reversal. The effect is captured by assuming $g(\gamma)$ becomes constant and equal to the value at the reversal strain. Irreversible damping does not change the results for strain-increasing deformations but can improve the accuracy of predictions for reversing strain histories. This is seen in data for polyisobutylene in **Figure 48** [10]. The over-prediction of the stresses by eqn (5.38) is corrected by assuming the value of $g(\gamma_l)$ is retained after the total strain becomes $\gamma_l + \gamma_2$.

Since networks formed by random crosslinking have a significant number of non-load bearing chain ends, Mullins softening in rubbers could have a similar anelastic origin. Harwood et al. [78] originally ascribed the effect to nonaffine displacement of the network junctions, in particular those near the end of chains. This idea can be tested by experiments on end-linked networks, which have a negligible chain end concentration. However, results on endlinked PTHF [80] show similar discrepancies with K-BKZ calculations for reversing strains (see **Figure 49**); thus, the problem cannot be the chain ends *per se*. The accuracy of the calculation can be improved by invoking irreversible damping but the agreement remains imperfect. Such experiments on endlinked, monodisperse chains also rule out a distribution of network strand lengths as the source of Mullins softening, although such network inhomogeneity can make the response less affine.

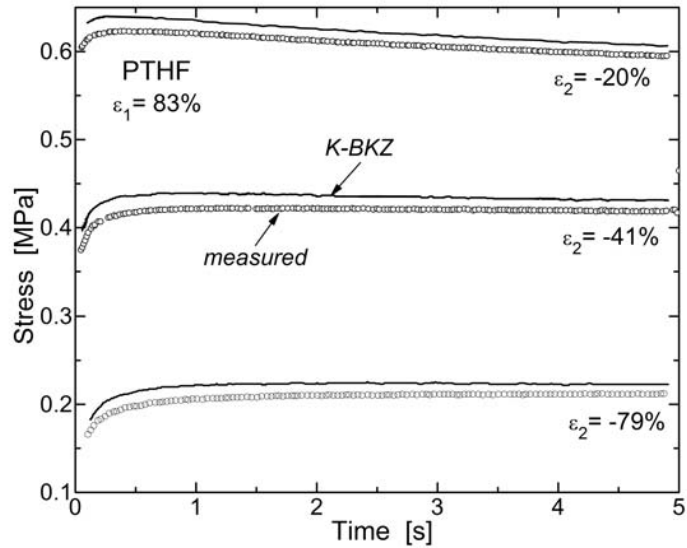

Figure 17. Experimental (circles) and calculated (solid lines) shear stresses measured in end-linked PTHF after double step shear strains of the indicated magnitudes and sign. The second step was applied at $t_l = 1$ s after the initial strain. Data from ref. [80].

## 5.3.2. Birefringence during stress-free recovery

During the recovery of a network or well-entangled polymer after the load is released, such as during recovery from creep flow, there is no external stress; nevertheless, measurable birefringence can be detected since there remains net orientation. This stress-free birefringence was first reported for natural rubber undergoing recovery from uniaxial stretching (Figure 50) [90]. As the applied stress was increased incrementally, proportionality was maintained between $\Delta n$ and $\hat{\sigma}$, even though the mechanical response became highly nonlinear. Upon removal of the load, a small birefringence persists, subsequently decaying slowly to zero as the chain segments assume an isotropic configuration. These results emphasize that birefringence arises due to segment orientation, which is not necessarily proportional to the stress.

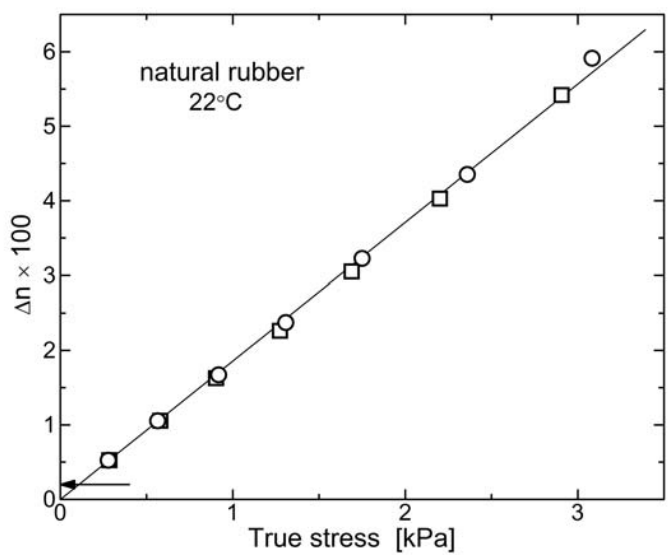

**Figure 18.** Birefringence from uncrosslinked NR under uniaxial tension. Measurements were made upon increment of the stress and after some subsequent creep. The arrow indicates the magnitude of $\Delta n$ measured about 2 min after removal of all stress. Data from ref. [90].

Creep-recovery experiments on rubber have shown even more anomalous behavior. After removal of the load, the residual birefringence has a sign opposite to that under load, signifying a *reversal* in the orientation. This anomaly was seen in 1,4-polybutadiene networks (**Figure 51**) [27], in linear polyisobutylene (**Figure 52**) [91], as well as in glassy PVC [92]. The change in sign of $\Delta n$ implies that the chain segments net orientation is opposite both to their orientation under stress and to the macroscopic orientation during recovery.

There is no diminution in the magnitude of the recovery $\Delta n$ when the molecular weight of the polymer is varied by more than an order of magnitude [91]. This argues against the above hypothesis for the anomalous relaxation of polymers melts subjected to strain reversals, rapid response of chain ends upon removal of stress, since the chain ends concentration varies inversely with $M_n$. It has been suggested [37,93] that when the stress is removed prior to attainment of mechanical equilibrium, the recovery is governed by a competition between unrelaxed chains and those having equilibrated to the imposed strain. A similar idea was invoked to describe the recoil following partial stress relaxation of polymer melts [94]. Thus, the change in sign of the birefringence upon a strain reversal may reflect a delicate balance within the distribution of chain orientations during deformation, prior to attainment of steady state. It would be interesting to measure the birefringence during recovery from steady state creep flow. Nevertheless, a change in sign of the net birefringence is not easily reconciled in this manner and it represents another aspect of the behavior of polymers subjected to reversing deformations whose origin remains to be understood.

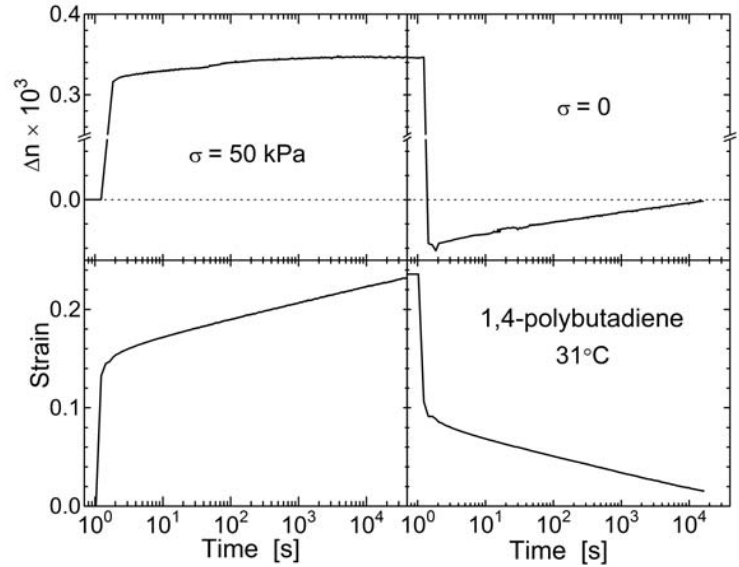

**Figure 19.** Shear strain and associated birefringence of cross-linked PBD during creep and recovery. The birefringence becomes measurably invariant at $10^4$ s, although the compliance never reaches steady state. The sign of $\Delta n$ changes upon removal of the load. Data from ref. [27].

## 5.4 Empirical rules for nonlinear flow

The zero-shear-rate limiting viscosity describes the shearing of unperturbed, fully entangled chains and $\eta_0$ can be measured either in a dynamic experiment or during steady flow. At higher rates of flow, the viscosity of high molecular polymers decreases, a result of orientation and disentanglement of the chains. This non-Newtonian behavior is also observed in dynamic measurements, even though the amplitude of the oscillation is small and thus the strains are negligible. Shear-thinning behavior is governed by the shear rate, and as first discovered by Cox and Merz in experiments on polystyrene [95], for equal values of the shear rate and frequency, the steady state shear and complex dynamic viscosities are essentially the same

$$\eta^*(\omega) = \eta(\dot{\gamma}) \qquad (\omega = \dot{\gamma}) \qquad (5.39)$$

This relation is strictly empirical, although efforts have been made to derive a theoretical basis [19,96,97] or at least determine the form of the relaxation function necessary for the K-BKZ equation (eqn (5.4)) to yield the Cox-Merz rule [98,99].

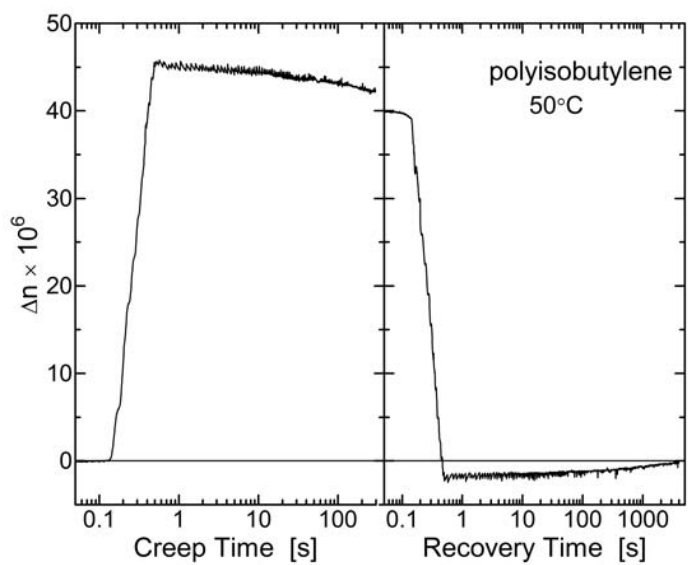

**Figure 20.** Strain and birefringence for PIB ($M_v = 9 \times 10^5$ g/mol). Data below *ca.* 1 s reflect the instrumental rise time. The decrease in *Δn* during creep is caused by the reduction in stress due to the changing cross-sectional area; the stress-optical coefficient actually increases slightly with strain. The birefringence has an opposite sign during the stress-free recovery. Data from ref. [91].

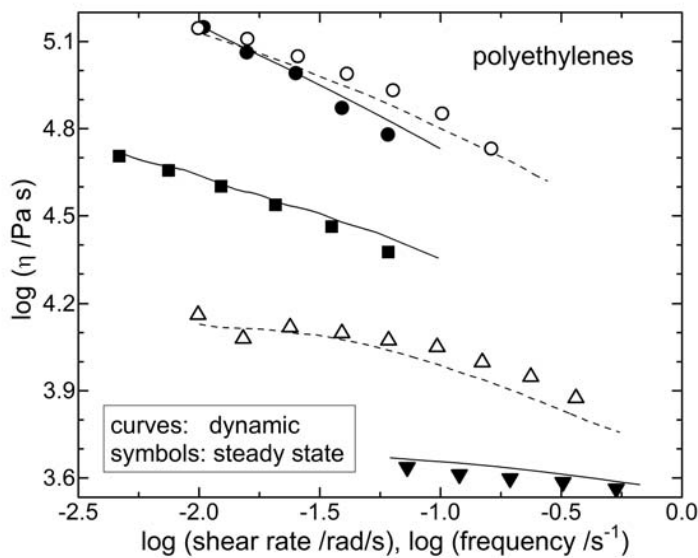

**Figure 21.** Comparison of $\eta^*(\omega)$ (symbols) and $\eta(\dot\gamma)$ (lines) for high density polyethylenes (HDPE, composed of primarily linear chains; solid symbols;) and low density polyethylenes (LDPE, which has branching on about 2% of its carbon atoms; open symbols). Different symbols correspond to different $M_w$, in the range from 100 – 140 kg/mol with substantial polydispersities ($\geq 4$). Temperature for HDPE is 190ºC and for LDPE = 175ºC. Data from ref. [103].

For linear polymers eqn (5.39) holds fairly well, especially for monodisperse, highly flexible chains [100,101]. A series of tests at 11 laboratories found that for neat polymers at modest shear rates the agreement was within the scatter of the experimental data [102]. At higher strain rates that induce chain stretching, dynamic viscosities are usually somewhat larger than the steady state values [98,103,104,105,106,107]. Some results are shown in **Figure 53** for various polyethylenes with broad molecular weight distributions [103]; the difference between $\eta(\dot\gamma)$ and $\eta^*(\omega)$ is about 10%. In **Figure 54** are viscosity data for neat polypropylene and a dynamically vulcanized blend of PP with EPDM [108]. Dynamic vulcanization refers to thermoplastic elastomers in which dispersed rubber particles are sulfur-crosslinked during mixing with the saturated polymer component [109]. Unlike the neat PP, the blend shows significant departures from eqn (5.39). Presumably flow orients and deforms the dispersed phase, reducing the viscosity from the value obtained in a small strain dynamic measurement. Similarly, in filled systems dynamic viscosities tend to exceed steady state values (**Figure 55**) [102,110].

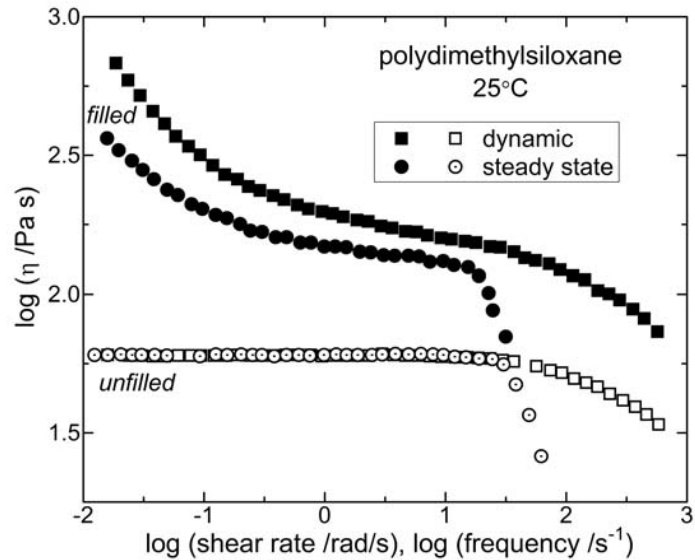

**Figure 22.** Comparison of $\eta^*(\omega)$ (squares) and $\eta(\dot\gamma)$ (circles) for neat PDMS (open symbols) and hydroxyl-terminated PDMS containing 12.9% by volume calcium carbonate (solid symbols) ($M_w = 1.3 \times 10^5$ g/mol). Only the former conforms to the Cox-Merz rule. The steep decrease in the steady state viscosity at shear rates > 20 rad/s is due to flow instabilities as the test specimen begins to fracture. Data from ref. [110].

Two empirical relationships due to Laun [111] relate the dynamic moduli to the first normal stress coefficient measured during steady state flow

$$\psi_1^0(\dot\gamma) = 2(G'/\omega^2)\left[1+(G'/G'')^2\right]^{0.7} \qquad (\dot\gamma = \omega) \qquad (5.40)$$

and to the steady state recoverable strain

$$\gamma_r(\dot\gamma) = (G'/G'')\left[1+(G'/G'')^2\right]^{1.5} \qquad (\dot\gamma = \omega) \qquad (5.41)$$

The former equation is more commonly used, with typical results shown in **Figure 56** for polyethylene and polypropylene [111] and in **Figure 57** for two polystyrenes with different molecular weight distributions [112]. For the latter the deviation of eqn (5.40) from the steady state values of $\psi_1^0$ is substantial. Often better agreement with experiments can be achieved by adjusting the exponent from its original value of 0.7; however, when employed as a fitting curve, eqn (5.40) loses its predictive capability. The terminal dynamic moduli are

$$\lim_{\omega \to 0} \frac{G'(\omega)}{\omega^2} = \eta_0^2 J_s^0 \quad ; \quad \lim_{\omega \to 0} \frac{G''(\omega)}{\omega} = \eta_0 \qquad (5.42)$$

whereby eqn (5.40) reduces at low shear rates to the relation of Coleman and Markovitz [113]

$$\psi_1^0 \equiv \lim_{\dot{\gamma} \to \infty} \frac{N_1(\dot{\gamma})}{\dot{\gamma}^2} = 2J_s^0 \eta_0^2 \qquad (5.43)$$

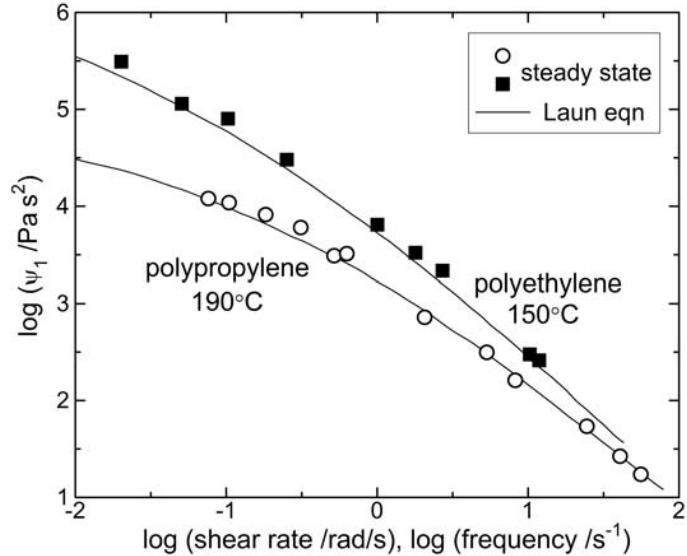

**Figure 23.** Experimental (symbols) and calculated (eqn (5.40); solid) first normal stress coefficient for polypropylene (circles) and HDPE (squares). Data from ref. [111].

The strain recovered by rubber subsequent to forming operations that squeeze the material (such as extrusion and calendaring) is known as die swell, $\hat{D}$, defined as the ratio of the final and initial dimensions transverse to the flow. Die swell is governed by the normal force[*], although only for steady state conditions is there a fundamental connection between $\hat{D}$ and strain rate. Two equations relate die swell to the first normal force [114]

---

[*] Although die swell in polymers becomes marked at strain rates associated with non-Newtonian viscosities, even Newtonian fluids lacking elasticity exhibit die swell, with $\hat{D} \sim 1.2$ under isothermal conditions. This is caused by changes in the velocity profile.

$$N_1 = 2\sigma\left(2\hat{D}^6 - 2\right)^{1/2} \tag{5.44}$$

and [115]

$$N_1 = 2\sigma\left(2\hat{D}^4 - \hat{D}^{-2}\right) \tag{5.45}$$

These can be used with the Lodge-Meissner equation (eqn (5.8)) to connect the die swell to the shear strain. Or, in combination with Laun's rule, eqns (5.44) or (5.45) provide estimates of $\hat{D}$ from dynamic mechanical measurements. Some results are shown in **Figure 58** for a filled EPDM rubber [116]. There is rough agreement with experimentally measured $\hat{D}$, which is satisfactory given the approximate nature of the derivation of eqn (5.44) or (5.45) [114,115].

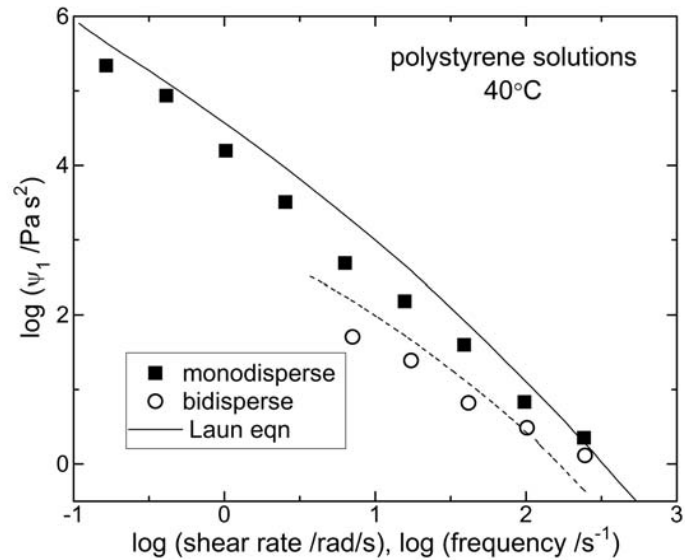

**Figure 24.** Experimental (symbols) and calculated (eqn (5.40); solid and dashed lines) first normal stress coefficient for monodisperse ($M_w = 2\times10^6$ g/mol; $P = 1.09$) and bidisperse ($M_w = 7\times10^5$ g/mol blended with 8% $M_w = 2\times10^6$ g/mol) polystyrene 20% solutions. Data from ref. [112].

An empirical equation was suggested by Gleissle [117] to estimate the steady state viscosity and its dependence on shear rate from the stress growth during startup at a slow (Newtonian) shear rate. The transient viscosity, $\eta^+(t)$, during the approach to steady-state is related to $\eta(\dot{\gamma})$ as

$$\lim_{\dot{\gamma}\to 0}\eta^+(t,\dot{\gamma}) = \eta(\dot{\gamma}) \qquad (t^{-1} = \dot{\gamma}) \tag{5.46}$$

Although the agreement with experimental data is not quantitative, this equation is quite useful because it provides estimates of viscosities over a range of shear rates from a single experiment. **Figure 59** compares $\eta^+(t)$ to both $\eta(\dot{\gamma})$ and $\eta^*(\omega)$ for linear and branched PIB [107]. Results for a PDMS are displayed in **Figure 60** [118]. Others assessments of the Gleissle relation have been reported for PIB [117] and polyethylene [111,119]. Gleissle [117] proposed an expression for the normal stress, analogous to eqn (5.46); however, in that case the relationship between reciprocal time and shear rate requires a numerical factor, which must be determined empirically for a given substance. This requires knowledge of $\eta(\dot{\gamma})$, obviating the purpose of the approximation, to circumvent such measurements.

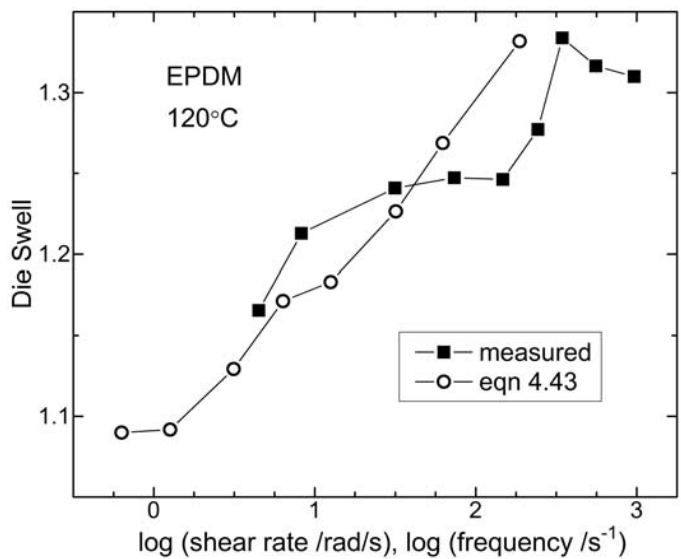

**Figure 25.** Experimentally measured $\hat{D}$ (solid square) for carbon black reinforced EPDM compound sheared in a capillary rheometer, along with the values (open circles) calculated from dynamic measurements using eqns. (5.40) and (5.44). Data from ref. [116].

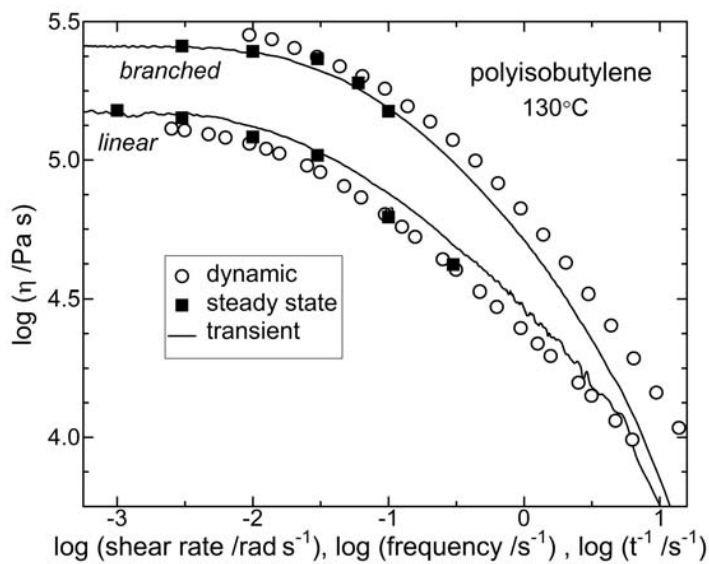

**Figure 26.** Viscosities determined from dynamic (open circles), steady state (solid squares), and the transient (eqn (5.46); line) shear stress measurements on a linear ($M_w = 5.0 \times 10^5$ g/mol) and a highly branched ($M_w = 1.08 \times 10^6$ g/mol) PIB. Data from ref. [107].

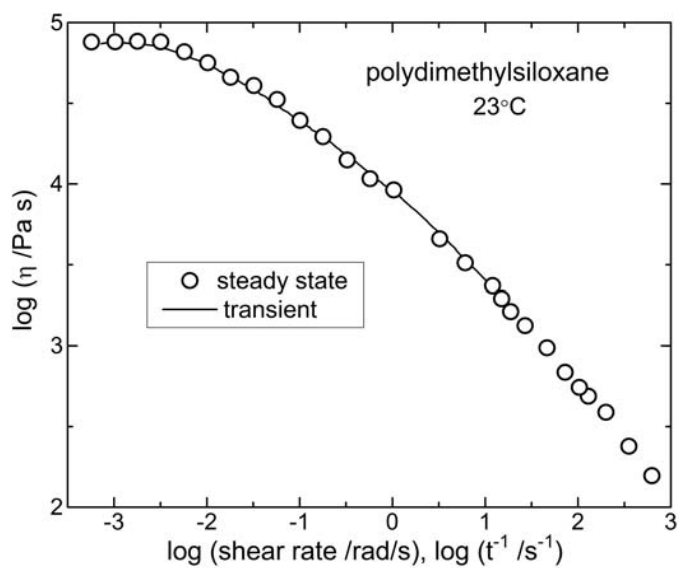

**Figure 27.** Viscosities determined from steady state (symbols) and the transient (eqn (5.46); line) shear stress measurements on a branched PDMS ($M_w = 4.7 \times 10^5$ g/mol). Data from ref. [118].

## 5.5 Payne effect

Utilization of rubber often relies on the incorporation of small-particle fillers, which improve processability by reducing die swell and generally increase both the stiffness and strength of the elastomer. Since most fillers are less expensive than the polymer, they also reduce costs. Rubber mixed with carbon black is the most widely used polymeric composite. The use of carbon black as a pigment dates to 4000 BC and as a reinforcing agent for rubber since the early 20[th] century. Methods to incorporate nanometer-sized fillers, the distribution and dispersion of the particles, and their effect on properties are aspects of the technology that have been widely reviewed [120,121,122,123,124,125,126]. Interest has been rekindled recently as part of the burgeoning attention to nanoscience and technology, with carbon nanotubes [127,128], silica nanoparticles [129,130], nanoclay [131,132,133], graphene [134,135 ], and diamond nanoparticles [136] investigated as potential fillers for polymers.

The effect of reinforcing fillers[*] on the modulus is very strain dependent. At all strains, stress amplification in the vicinity of the inextensible particles [137,138,139] increases the modulus. This hydrodynamic effect is enhanced by the occlusion of rubber at the particle interface [140,141]. At higher strains various mechanisms, specific to the material, become operative, including detachment of chains from the filler [142], orientation of the dispersed particles [143], and microscopic crack deviation around the filler phase [144]. All contribute to reinforcement and increase the nonlinearity of the mechanical behavior of filled rubber.

---

[*] Fillers that do not "reinforce" the polymer are referred to as extenders, since their chief purpose is to increase the quantity of material and thus lower its cost. Extenders can increase the modulus by amplifying local stresses; however, they are usually relatively large and interact negligibly with the polymer chains. The result is that the modulus enhancement is modest and failure properties such as elongation and fatigue life usually become worse.

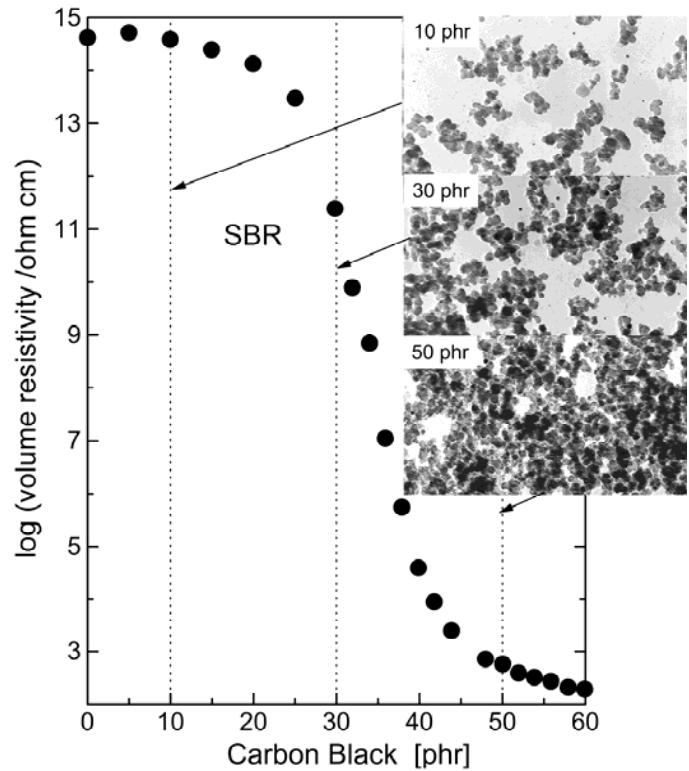

**Figure 28.** Transmission electron micrograph images of SBR containing the indicated quantity of N330 carbon black. The corresponding electrical resistivity behavior is indicated in the graph. Data from ref. [147].

Although flocculation of dispersed particles is common to virtually all fillers, agglomeration of carbon black has received the most attention. Carbon black exists as aggregates, formed during the manufacturing process and ranging in size from about 10 to 100 nm. When the filler concentration exceeds the percolation threshold, interaction among the aggregates leads to formation of a continuous network structure. This filler network causes an enormous increase in electrical conductivity [145,146,147] (**Figure 61**), the appearance of yield stresses (thixotropy), and an inverse variation with strain amplitude of the dynamic modulus [148,149,150]. The latter is due to the breakup of the flocculated structure as the mechanical stress overcomes the van der Waals forces responsible for aggregate bonding [142,143,144,151]. This breakup of the filler network dissipates energy [152] and thus can be a significant contributor to, for example, the fuel consumption of motor vehicle tires [153].

In a typical dynamic experiment, harmonic shear deformation is imposed at an excitation frequency $\omega$

$$\gamma^*(t) = \gamma_0(t)\exp(i\omega t) \tag{5.47}$$

After steady state is attained (requiring a duration exceeding the relevant relaxation time, with a few cycles usually sufficient), the linear response is also a single harmonic function at the same frequency, albeit with a phase shift due to any loss

$$\sigma^*(t) = \sigma_0(t)\exp(i[\omega t + \delta])$$ (5.48)

(This complex representation is a convenience; the actual motion is sinusoidal.) In terms of the dynamic moduli,

$$G^* = \sigma^* / \gamma^*$$ (5.49)

$$G' = (\sigma_0 / \gamma_0)\cos(\delta)$$ (5.50)

and

$$G' = (\sigma_0 / \gamma_0)\sin(\delta)$$ (5.51)

If the response depends either on the amplitude, $\gamma_0$, or frequency, it is no longer described by eqn (5.48). However, an arbitrary periodic function can be expressed as a sum of harmonics, so that departures from linearity give rise to intensity at frequencies other than the fundamental $\omega$. If the nonlinearity is approximated as a power series in the strain or strain rate, this superposition has the form [154]

$$\sigma(t) = \sigma_0\cos(\omega t) + \sigma_3\cos(3\omega t) + \sigma_5\cos(5\omega t) + ...$$ (5.52)

That is, the nonlinearities appear as odd harmonics in the measured response.[*]

Although disruption of the agglomerated particles introduces strong nonlinearities into the dynamic mechanical behavior, the response of filled rubber at any given strain remains linear; thus, the sinusoidal response lacks higher order harmonics [155,156,157]. A steady-state degree of agglomeration is maintained, as determined by the maximum strain. At low strains the modulus is constant, as the filler network remains intact; higher strain amplitudes reduce the stiffness, with the appearance of a maximum in the loss modulus (**Figure 62**) [149]). This characteristic behavior is known as the Payne effect [158][†]159,. The same phenomenon is seen in colloidal suspensions and other thixotropic fluids [161,162]. The Payne effect is not accompanied by any nonlinearity during sinusoidal deformation at fixed amplitude, because the network does not recover on the time scale of the dynamic perturbation. This very slow recovery of the filler network is illustrated in **Figure 63**, showing the gradual recovery of the storage modulus and dc-electrical conductivity upon reduction in the strain amplitude.

---

[*] These higher harmonics resulting from material nonlinearity or experimental distortions are distinct from the application of multiple input frequencies in order to simultaneously measure the linear response at several frequencies ("Fourier transform rheometry") [**Error! Bookmark not defined.**]. The latter is an efficient experimental method, especially useful for transient structures; however, signal to noise issues are a consideration given the need to remain in the linear regime [**Error! Bookmark not defined.**].

[†] The dependence on amplitude of the dynamic response unique to filled rubber was noted earlier by Gehman et al. [159] and by Fletcher and Gent [160].

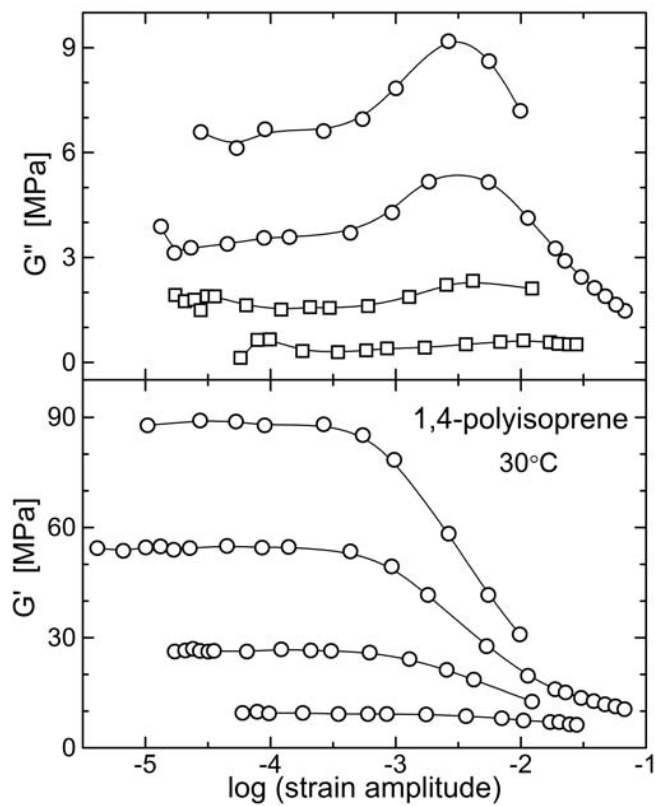

**Figure 29.** Dynamic storage (lower) and loss modulus (upper) versus strain for PI at 0.1 Hz and 30°C. At low strains the modulus is constant, with higher amplitudes disrupting the filler network. Data from ref. [149].

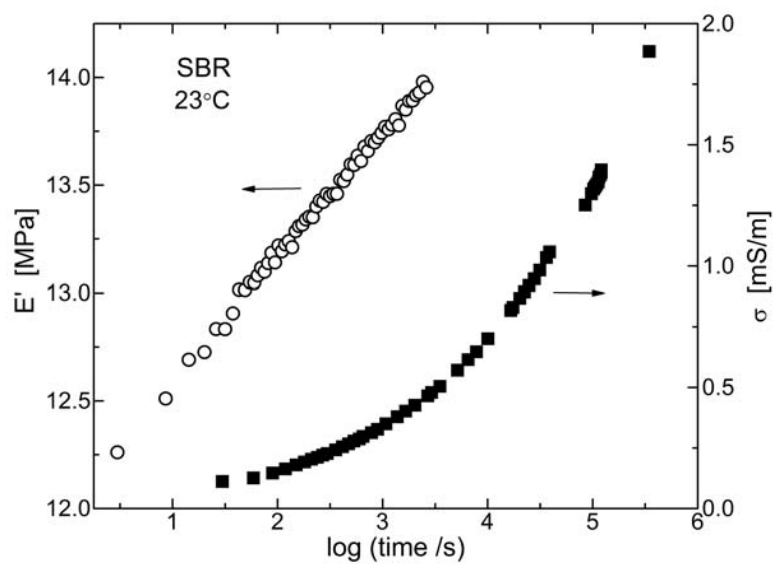

**Figure 30.** Low strain storage modulus (open circles) measured at 40 Hz for an SBR network with 82.5 phr N339 carbon black and 62.5 phr oil, following a static strain of 1.9%; the corresponding DC-conductivity (filled squares) is also shown. The values of $E'$ (= 15.0 MPa) and $\sigma_{DC}$ (= $4.2 \times 10^{-3}$ S/m) prior to the static strain are substantially larger than attained after recovery (which extended to four days for the electrical measurement).

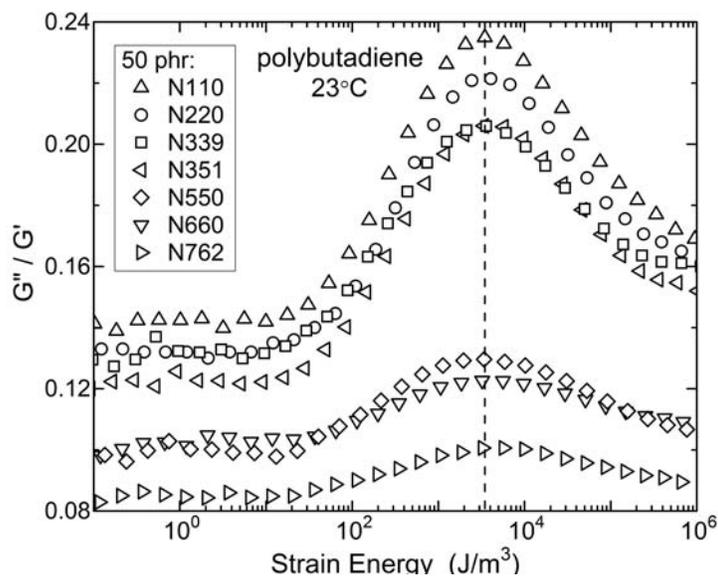

Figure 31. Loss tangent *versus* strain energy for polybutadiene with the indicated carbon black. The aggregate size of the carbon black varies from 54 nm (N110) to 188 nm (N762), the first two digits being a rough measure of the primary particle size (e.g., ~55 nm for N550, which has an aggregate size of 139 nm). The vertical dashed line denotes the characteristic strain energy, 3.5 kJ/m3, which is independent of filler size. Dynamic data were obtained at 0.5 Hz at shear strain amplitudes from $10^{-4}$ to 0.3 Data from ref. [165].

The Payne effect becomes evident at strain amplitudes in the range 0.1 – 1%. Interestingly, the disruption of the particle contacts transpires at a characteristic strain energy that is independent of the size or concentration of the particles [161,163,164,165]. This is illustrated in **Figure 64** showing the superpositioning of the loss tangent curves for SBR with a fixed concentration of varying size of carbon black particles [165]. The Payne effect is not specific to filled polymers, requiring only a flocculated particulate structure within a fluid medium; thus, oil containing silica particles shows the Payne effect, with strain energy again as the control parameter for the break up of the filler structure **Figure 65** [163].

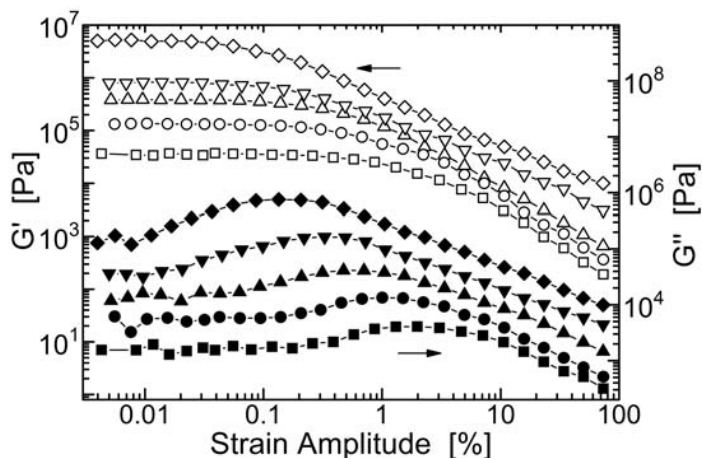

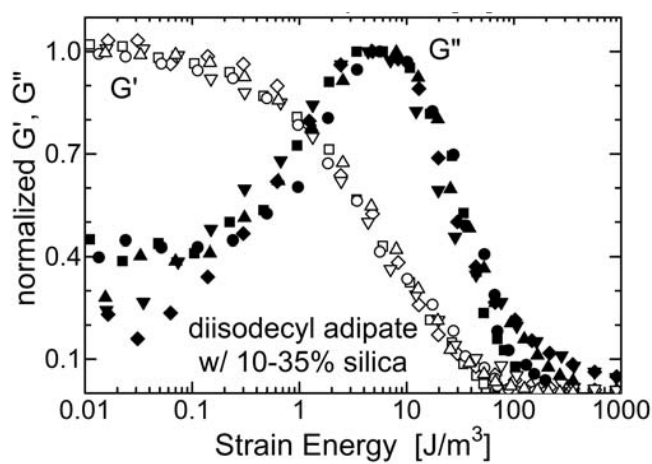

**Figure 32.** (upper) Dynamic storage (open symbols) and loss (filled symbols) moduli at 30ºC and 0.5 Hz *versus* strain amplitude for the diisodecyl adipate containing 0.10 (squares), 0.14 (circles), 0.20 (triangles), 0.30 (down triangles), and 0.35 (diamonds) volume fraction silica particles Data from ref. [164]. (lower) Same data with the storage modulus normalized by the low strain plateau value and the loss modulus by the maximum in G". Data from ref. [163].